\documentclass[aps,twocolumn,superscriptaddress,floatfix,nofootinbib]{revtex4-1}
\usepackage{graphicx,amsmath,amssymb,verbatim,color}
\usepackage{booktabs}
\usepackage{comment}
\usepackage{soul}
\usepackage{setspace}
\usepackage[dvipsnames]{xcolor}
\usepackage{bm}
\usepackage[utf8]{inputenc}
\usepackage[colorlinks=true,citecolor=blue,linkcolor=blue,urlcolor=blue, backref=false,pdfborder={0 0 0}]{hyperref}
\usepackage{float}
\usepackage{multirow}
\usepackage{dcolumn}

\newcommand{\Planck}{\textit{Planck}}
\newcommand{\LCDM}{$\Lambda$CDM}
\newcommand{\Plik}{\textit{Plik}}
\newcommand{\CamSpec}{\textit{CamSpec}}

\newcommand{\Hunit}{{\rm km}\ {\rm s}^{-1}\rm{Mpc}^{-1}}

\begin{document}

\pagenumbering{arabic}

\title{Improved Planck constraints on axion-like early dark energy as a resolution of the Hubble tension}

\author{George Efstathiou}
\email{gpe@ast.cam.ac.uk}
\affiliation{Kavli Institute for Cosmology \& Institute of Astronomy, Madingley Road, Cambridge, CB2 0HA, UK.}

\author{Erik Rosenberg}
\email{erik.rosenberg@manchester.ac.uk}
\affiliation{Jodrell Bank Centre for Astrophysics, School of Physics \& Astronomy, University of Manchester, Oxford Road, Manchester, M13 9PL, UK.}
\affiliation{Kavli Institute for Cosmology \& Institute of Astronomy, Madingley Road, Cambridge, CB2 0HA, UK.}
\author{Vivian Poulin}
\email{vivian.poulin@umontpellier.fr}
\affiliation{Laboratoire Univers \& Particules de Montpellier, CNRS \& Université de
Montpellier (UMR-5299), 34095 Montpellier, France.}

\date{\today} 
\begin{abstract}

Axion-like early dark energy (EDE) as an extension to \LCDM\ has been proposed as a possible solution to the 'Hubble tension'. We revisit this model using a new cosmic microwave background (CMB) temperature and polarization likelihood constructed from the \Planck\ NPIPE\protect\footnote{\underline{N}ational Research Scientific Computing Center \underline{pipe}line, described in \cite{Planck:2020olo} and available from the Planck Legacy Archive, https://pla.esac.int.} data release. In a Bayesian analysis, we
find that the maximum fractional contribution of EDE to the total energy density is $f_{\rm EDE} < 0.061$ (without SH0ES) over the redshift range $z\in[10^3,10^4]$ and that the
Hubble constant is constrained to  lie within the range $ 66.9 < H_0 < 69.5 \ \Hunit$ (both at 95 \% C.L.).  The data therefore favour a model close to \LCDM, leaving a residual tension of $3.7\sigma$ with the SH$0$ES  Cepheid-based  measurement of $H_0$. A comparison with the likelihood profile shows that our conclusions are robust to prior-volume effects. Our new 
CMB likelihood provides no evidence in favour of a significant  EDE component. 
\end{abstract}
\maketitle

Recent improvements in the determination of the Hubble constant $H_0$ have led to a potential crisis in cosmology. Assuming the standard six-parameter \LCDM\  model, CMB data from \Planck\ predicts\footnote{Unless otherwise stated, in this paper two-sided constraints are quoted at 1$\sigma$, while one-sided constraints are quoted at $2\sigma$.}
$H_0=67.5\pm0.5\ \Hunit$ \cite{ParamsIII}. However,  direct measurement of $H_0$ via Cepheid-calibrated Type Ia supernovae (SN1a) by the SH$0$ES collaboration yields $H_0=73\pm1 \ \Hunit$, in apparent $5\sigma$ tension with the \LCDM\  prediction \cite{Riess:2021jrx}. Although not all measurements of the Hubble constant are in strong tension with the \Planck\ \LCDM\ prediction \cite{Freedman:2020dne,Abdalla:2022yfr}, increasingly stringent tests of the Cepheid-based 
distance scale have failed to reveal evidence of systematics in the SH$0$ES data
 \cite{Riess:2021jrx, Yuan:2022, Riess:2022,Riess:2023}. It is therefore important to carefully consider the possibility that the `Hubble tension' is caused by physics beyond \LCDM.

Among many models proposed to resolve the Hubble tension \cite{Schoneberg:2021qvd,DiValentino:2021izs,Abdalla:2022yfr}, Early Dark Energy (EDE) has emerged as  one of the most plausible \cite{Poulin:2018cxd,Knox:2019rjx,Kamionkowski:2022pkx,Poulin:2023lkg}, since it is able to reduce the Hubble tension to about the $2\sigma$ level while maintaining a good fit to 2018 \Planck\ CMB power spectra, Baryonic Acoustic Oscillations (BAO) and SN1a data. An EDE solution to the Hubble tension would also have important implications for our understanding of inflationary cosmology, since acceptable fits to the CMB power spectra require a nearly scale-invariant spectrum of curvature perturbations. Thus EDE  affects the interpretation of the dynamics of inflation and  of   gravitational wave 
upper limits from observations of CMB polarization on large angular scales \cite{Kallosh:2021, Ade:2021}. 

A consensus on the observational constraints on EDE has yet to emerge. Bayesian analyses of the \Planck\ data tend to disfavor EDE when  the SH$0$ES measurements are excluded \cite{Hill:2020osr}. 
Nevertheless, it has been argued that these Bayesian analyses are affected by prior volume effects \cite{Murgia:2020ryi,Smith:2020rxx}.  Profile likelihood analyses based on the  \Planck\  \Plik\ temperature-polarization likelihood \cite{Planck:2019nip}
lead to weaker constraints and may even favor a non-zero EDE component with a larger $H_0$ at about $2\sigma$, mainly driven by the polarization spectra \cite{Herold:2021ksg,Herold:2022iib}. The recent ACT DR4 data have hinted at a $3\sigma$ preference for EDE over \LCDM\ \cite{Hill:2021yec,Poulin:2021bjr,Smith:2022hwi}, though this result is not supported by the recent 2018 SPT3G data  \cite{Smith:2023oop} or by \Planck\ temperature  measurements.

\begin{figure*}[t!]
    \centering
{\hspace{-0.4truein}\includegraphics[width=1.08\textwidth]{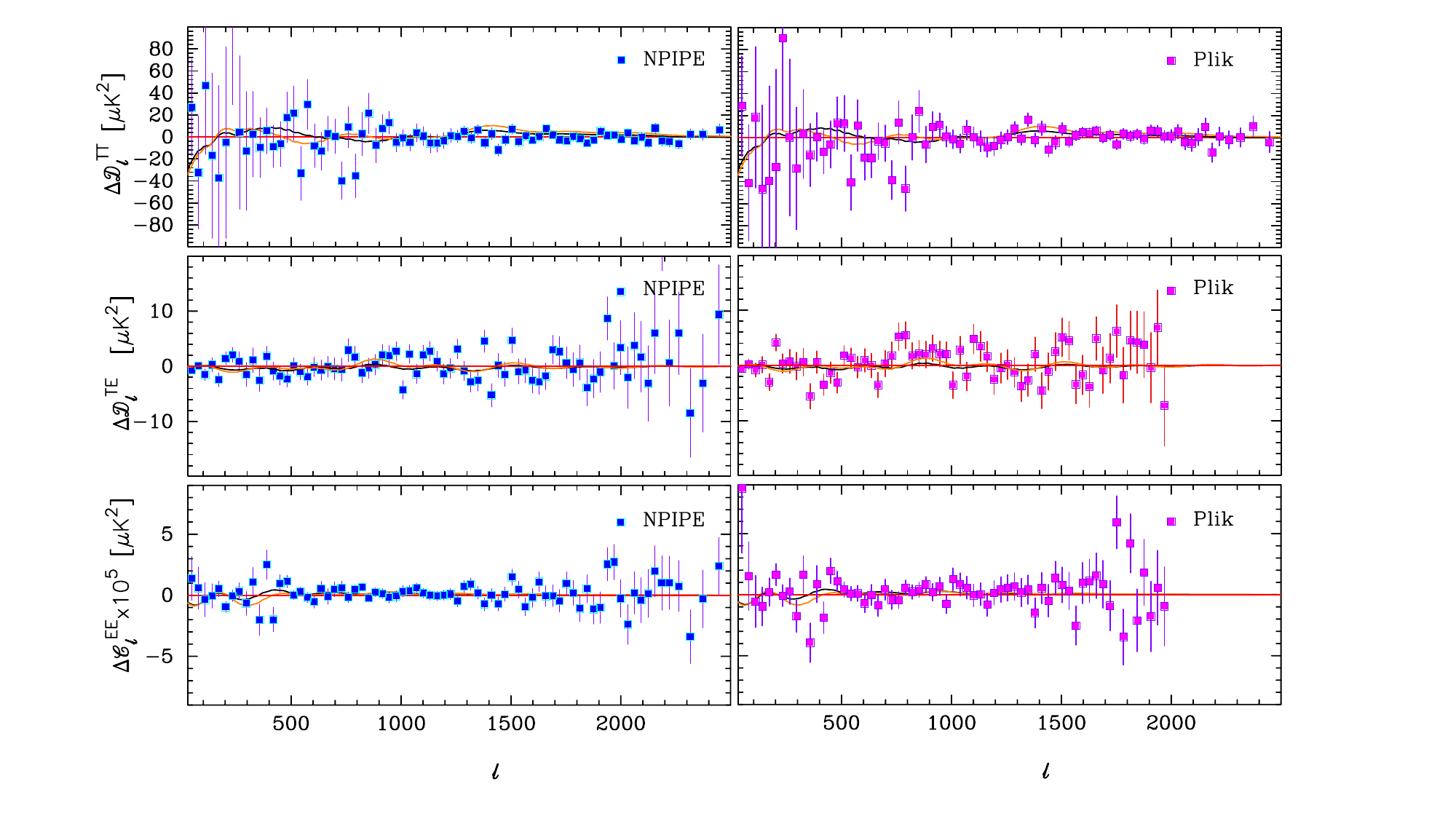}}
    \caption{Residuals of the TT, TE and EE spectra relative to the best fit \LCDM\ model to the NPIPE TTTEEE. Residuals for NPIPE spectra used in this {\it letter} are shown in the left hand panels. Residuals  for the \Plik\ spectra,  as used in the baseline  2018 \Planck\ TTTEEE likelihood, relative to the same cosmology are shown in the right hand panels. The lines show the residuals of the best fit EDE model,   summarized in Table \ref{tab:EDE_MCMC}, fitted to NPIPE+SHOES (black) and \Plik+SH0ES (orange). Note that the EE spectrum at $\ell > 2000$ is not used in the NPIPE likelihood.}
    \label{fig:ede_spectra}
\end{figure*}

Since the 2018 \Planck\ likelihood release \cite{Planck:2019nip}, it has become possible to construct more powerful  CMB likelihoods by using more sky area   at high frequencies \cite{Efstathiou:2021, Rosenberg:2022sdy}. In addition, the recent NPIPE  maps
\cite{Planck:2020olo} incorporate a number of improvements in the processing of time ordered
data,  leading to a significant reduction in the small scale noise compared to previous \Planck\ data releases.  In this {\it letter}, we use an updated \Planck\ likelihood  based on the work of Ref.~\cite{Rosenberg:2022sdy} to investigate  the viability of EDE as a solution to the Hubble tension and to assess the potential hints for EDE reported in some  earlier analyses of \Planck\ and ACT DR4 data.

We analyze the popular axion-like EDE model, specified by the modified axion potential \cite{Kamionkowski:2014zda,Poulin:2018cxd,Smith:2019ihp}
\begin{equation}\label{eq:potential}
    V(\theta) = m^2 f^2[1-\cos (\theta)]^3,
\end{equation} where $m$ represents the axion mass, $f$ the axion decay constant, and $\theta \equiv \phi/f$ is a re-normalized field variable defined such that $-\pi \leq \theta \leq \pi$.  This potential provides a EDE model with flexible phenomenology, and can be embedded in a string-theory framework \cite{McDonough:2022pku,Cicoli:2023qri}. For other EDE parametrizations, we refer to Refs.~\cite{Agrawal:2019lmo,Lin:2019qug,Alexander:2019rsc,Sakstein:2019fmf,Das:2020wfe,Niedermann:2019olb,Niedermann:2020dwg,Niedermann:2021vgd,Ye:2020btb,Berghaus:2019cls,Freese:2021rjq,Braglia:2020bym,Sabla:2021nfy,Sabla:2022xzj,Gomez-Valent:2021cbe,Moss:2021obd,Guendelman:2022cop,Karwal:2021vpk,McDonough:2021pdg,Wang:2022nap,Alexander:2022own,McDonough:2022pku,Nakagawa:2022knn,Gomez-Valent:2022bku,MohseniSadjadi:2022pfz,Kojima:2022fgo,Rudelius:2022gyu,Oikonomou:2020qah,
Tian:2021omz,Maziashvili:2021mbm}.

The 2018 \Planck\ data release (hereafter PR3) used the \Plik\ temperature-polarization power spectrum  likelihood as the baseline for cosmological parameter analysis \cite{ParamsIII}.  
The \Planck\ collaboration papers also reported results using an alternative likelihood, \CamSpec\ (which was used as the baseline in the first \Planck\ data release \cite{ParamsI}). These two likelihoods are almost identical in temperature and differ primarily in polarization. As described in \cite{ParamsIII}, they give very similar results for most cosmological models. Following PR3, Ref.~\cite{Efstathiou:2021} presented an extension of \CamSpec\ which, via  cleaning of the $143$ and $217$ GHz temperature maps using $545$ GHz maps
as a template of Galactic dust emission, allowed the use of 80\% of sky.  The $TE$ and $EE$ spectra used in \CamSpec\ are cleaned from polarized dust emission using $353$ GHz maps. Extensive tests of these cleaning procedures including tests as a function of frequency, detector combination and sky coverage
are presented in Ref.~\cite{Efstathiou:2021}. 

Ref.~\cite{Rosenberg:2022sdy} applied the \CamSpec\ methodology to the 2020  \Planck\ data release (PR4) NPIPE
maps. The NPIPE maps are produced for each of two detector groupings,  
$A$ and $B$,  which are processed independently. The NPIPE likelihood produced in Ref.~\cite{Rosenberg:2022sdy} used only $A\times B$ cross spectra, but in this paper we include $A\times A$ and $B \times B$ cross spectra for different frequencies since we have found no evidence for correlated noise in these spectra. This leads to an improvement in the signal-to-noise of the 
coadded TT and EE spectra. The other aspects of the likelihood, frequencies, multipole ranges, foreground treatment in temperature, and calibration parameters are as described in Ref.~\cite{Rosenberg:2022sdy} except for extending TE to $\ell = 2500$.  The best-fit six parameter \LCDM\ model determined from this likelihood is almost identical to the best-fit  model presented in Ref.~\cite{Rosenberg:2022sdy}. Residuals of the NPIPE  coadded $TT$, $TE$ and $EE$ spectra\footnote{The $TT$ spectra are corrected for the best-fit foreground model. In \CamSpec\ there are no foregrounds in the dust-corrected $TE$ and $EE$ spectra.} with respect to this best-fit model are plotted in Fig. \ref{fig:ede_spectra}. The residuals of the 2018 \Plik\ spectra used in the publicly available \Plik\ likelihood are plotted in the right hand panels of Fig. \ref{fig:ede_spectra}. As can be seen, the NPIPE spectra have substantially smaller residuals compared to \Plik. Significantly  (emphasised in Ref.~\cite{Efstathiou:2021}), as the statistical power of the \Planck\ temperature and polarization spectra have improved, they have all come into even closer agreement with the predictions of the base six parameter \LCDM\ cosmology, with no evidence for any significant anomalies. As a consequence,  we anticipate that  the NPIPE \Planck\ spectra are likely to disfavor EDE as a solution to the Hubble tension.

\begin{figure}[t!]
    \centering
    \includegraphics[width=\columnwidth]{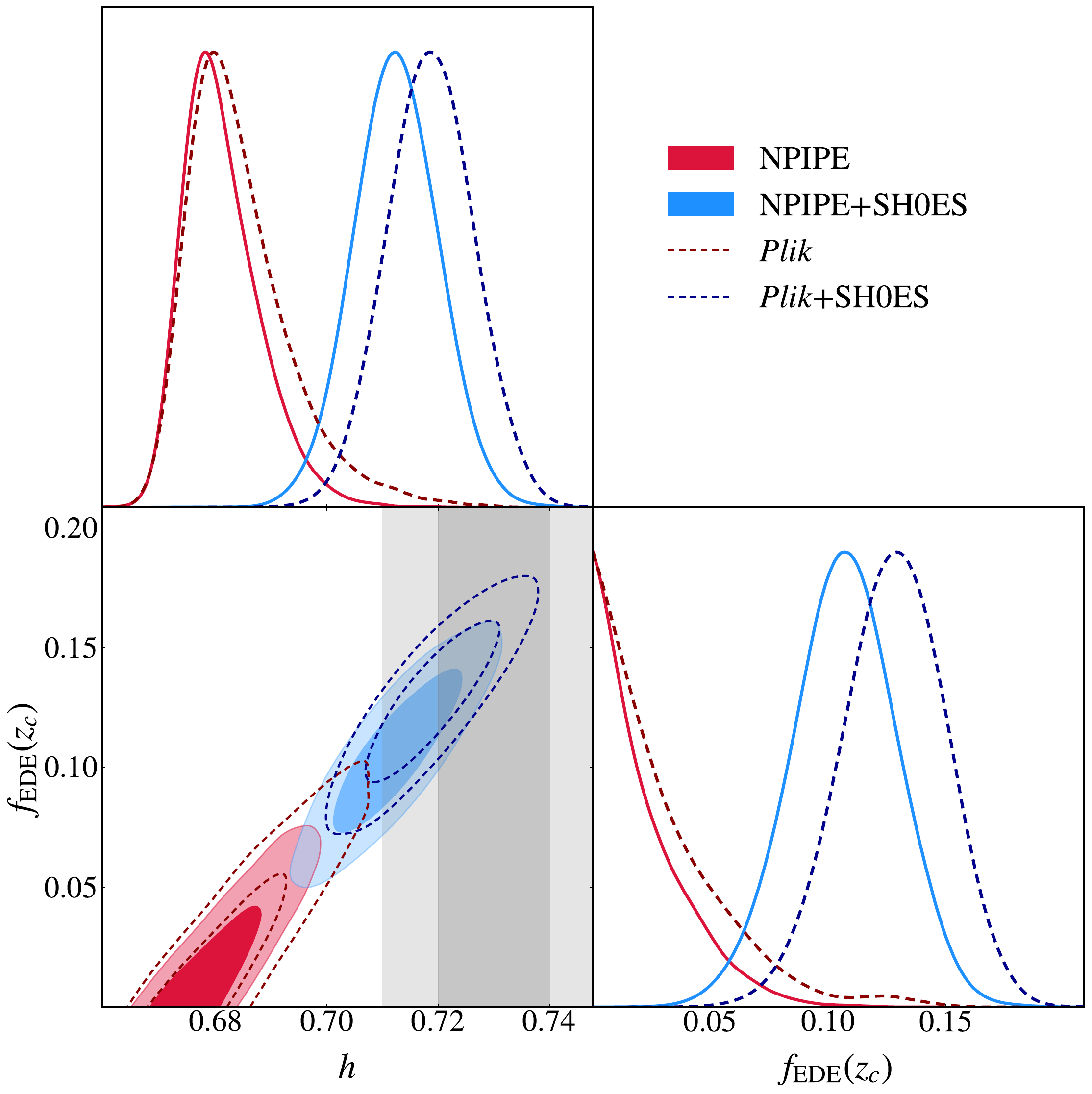}
    \caption{Posterior distributions  of $H_0$ and the EDE fraction $f_{\rm EDE}(z_c)$ 
    computed from analyses using the new NPIPE CMB likelihood compared to the 2018 \Plik\ likelihood. We report results with and without the SH0ES $M_b$ prior. All MCMC runs  include \Planck\  CMB lensing,  BAO/$f\sigma_8$ and Pantheon+ data as described in the text.}
    \label{fig:ede_MCMC}
\end{figure}

In addition to our  new \Planck\  likelihood, the baseline analysis in this paper includes the TT and EE likelihoods in the multipole
range $2 \le \ell < 30$ from \Planck~2018 \cite{Planck:2019nip} together with the `conservative' \Planck\ lensing likelihood \cite{Planck:2018vyg},  measurements of the BAO and redshift space distortions from the CMASS and LOWZ galaxy samples of BOSS DR12 at $z = 0.38$, $0.51$, and $0.61$ \cite{BOSS:2016wmc},   BAO measurements from 6dFGS at $z = 0.106$ and SDSS DR7 at $z = 0.15$  \cite{Beutler:2011hx,Ross:2014qpa} and  the Pantheon+ catalog of over 1600 SN1a which constrains the luminosity distance 
over the  redshift range $0.01 < z < 2.3$ \cite{Brout:2022vxf}. The SH0ES Cepheid calibration of the peak SN1a absolute magnitude is modelled as a Gaussian,  $M_b=-19.253 \pm
0.027$ \cite{Riess:2021jrx}. Consistency with SH0ES is assessed via the statistic \cite{Raveri:2019}
\begin{eqnarray}
& & Q_{\rm DMAP}=\sqrt{\Delta \chi^2} \qquad \qquad \qquad\nonumber \\
& &= \sqrt{\chi_{\rm tot}^2({\rm with ~SH0ES})-\chi_{\rm tot}^2({\rm without~SH0ES})}, \qquad 
\label{equ:Qdmap}
\end{eqnarray} where the two values of $\chi^2$ 
are computed at the maximum a posteriori points  with and without including the SH$0$ES $M_b$ prior\footnote{$Q_{\rm DMAP}$ gives an indication of the inconsistency betwen the SH$0$ES prior and our baseline likelihoods in terms of an effective 'number of $\sigma$'. Alternatively, one can compute a $p$-value from  $\Delta \chi^2$ assuming a $\chi^2$ distribution with 1 degree of freedom. }.

We run Markov-chain Monte Carlo (MCMC) using the Metropolis-Hasting algorithm implemented in {\sf MontePython-v3}\footnote{\url{https://github.com/brinckmann/montepython_public}} \citep{Audren:2012wb,Brinckmann:2018cvx} 
 interfaced with a modified version of {\sf CLASS}\footnote{\url{https://lesgourg.github.io/class_public/class.html}} \cite{Lesgourgues:2011re,Blas:2011rf}. We use large flat priors on $H_0$, the  baryon and cold dark matter energy density, $\omega_b$ and $\omega_{\rm cdm}$ respectively, the amplitude and tilt of the scalar perturbations $A_s$ and $n_s$ respectively, and the reionization optical depth $\tau_{\rm reio}$. For the EDE sector, we follow the approach in Ref.~\cite{Smith:2019ihp} and vary the critical redshift $z_c\in[10^3,10^4]$ after which the scalar-field starts rolling, the fractional contribution of EDE at that redshift $f_{\rm EDE}(z_c)\in[0,0.3]$ and the initial field value $\theta_i\in[0,\pi]$.  We model free-streaming neutrinos as two massless  and one massive neutrino with $m_\nu=0.06$ eV. We use {\sf Halofit} to estimate the non-linear matter clustering \cite{Smith:2002dz}.
We consider chains to be converged using the conventional Gelman-Rubin criterion $|R -1|\lesssim0.05$ \citep{Gelman:1992zz}. 
To analyze the chains and produce our figures we use the {\sf GetDist} package \cite{Lewis:2019xzd}, and we obtain the minimum $\chi^2$ values using the simulated annealing method described in the appendix of Ref.~\cite{Schoneberg:2021qvd}. In addition to MCMC, and to mitigate prior-volume effects that have been shown to affect posterior distributions \cite{Herold:2021ksg,Herold:2022iib,Poulin:2023lkg}, we present likelihood profiles of $H_0$ and $f_{\rm EDE}$, that are obtained using a methodology which will be presented in a forthcoming paper \cite{ProfInPrep}.

\begin{figure*}
    \centering
    \includegraphics[width=\columnwidth]{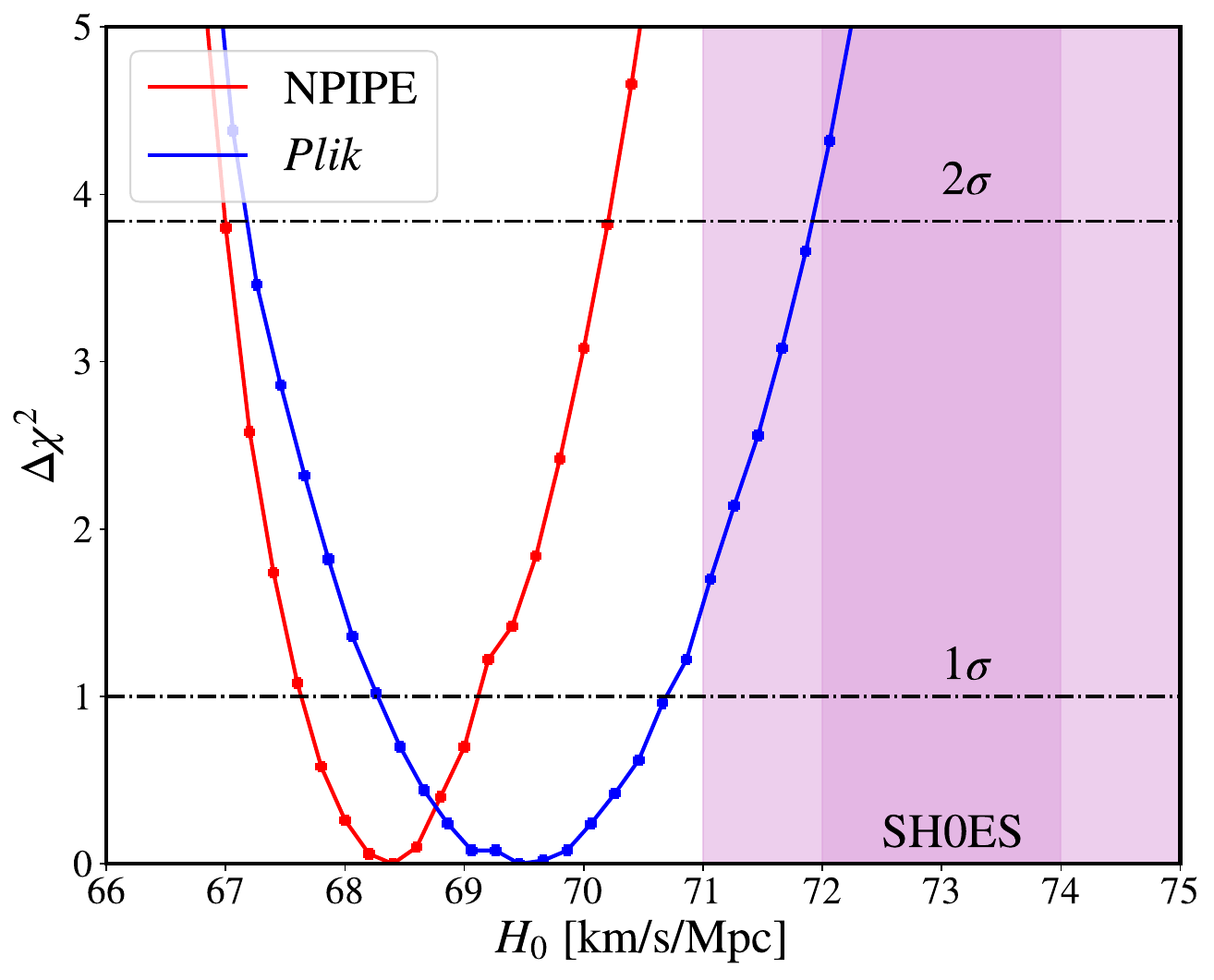}
    \includegraphics[width=\columnwidth]{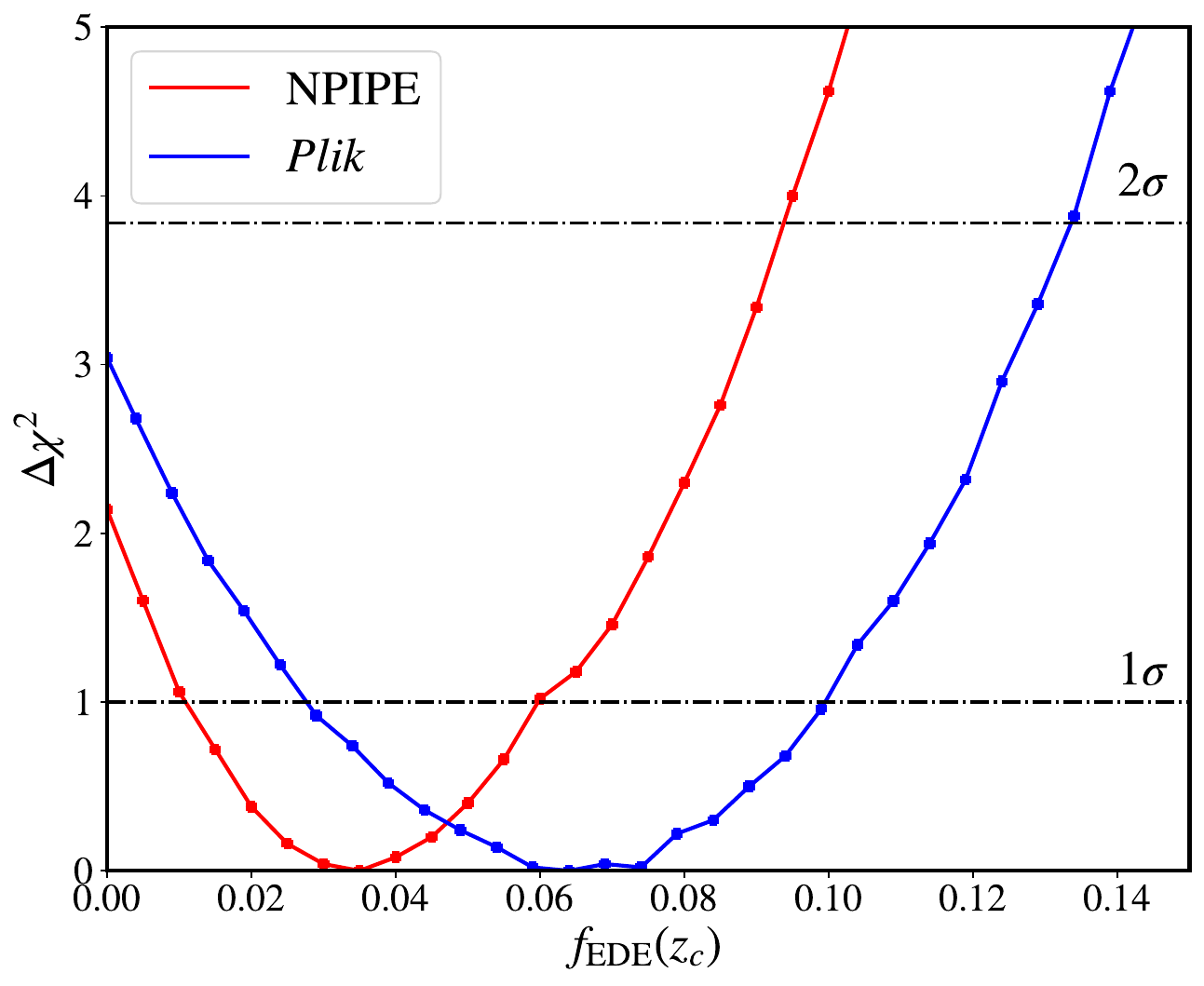}
        \setlength{\belowcaptionskip}{-10pt}
    \caption{Likelihood profile of $H_0$ and $f_{\rm EDE}$ reconstructed from our analyses
     using the new NPIPE likelihood compared to the 2018 \Plik\ likelihood.  All runs  
     include supplementary data  as described in the text.}

    \label{fig:ede_profile}

\end{figure*}

Our main results are shown in Fig.~\ref{fig:ede_MCMC} and reported in Table~\ref{tab:EDE_MCMC}. With the new  NPIPE likelihood we find $f_{\rm EDE}(z_c)<0.061$, with $h<0.696$  at $2\sigma$ (where $h$ is the Hubble constant in units of $100 \Hunit$). 
The addition of the SH0ES prior  raises the contribution of EDE to the $10\%$ level, with $h \simeq 0.712\pm0.008$, but this comes at the expense  of a subtantially worse fit to the NPIPE TTTEEE likelihood ($\Delta\chi^2 = +6.3)$ (see Table \ref{tab:chi2_Planck_EDE}). The solid black line in Fig.~\ref{fig:ede_spectra} shows the residuals of the best fit EDE model fitted to NPIPE and SH$0$ES. This fit gives a $\chi^2$ of $4.4$ for the SH$0$ES prior, and as a consequence 
of the poor fits to both the CMB and SH$0$ES data the $Q_{\rm DMAP}$ tension metric indicates
a $3.7\sigma$ discrepancy between SH$0$ES and our baseline datasets ($p$-value of $0.000239$).  These results are significantly stronger  than the  $\sim 2.6\sigma$ tension\footnote{The use of the updated Pantheon+ SN sample that raises $\Omega_m$ by $\sim 1 \sigma$ and the tighter $M_b$ determination by SH0ES
increases the tension by $\sim 0.5\sigma$ compared to previous results \cite{Simon:2023hlp}.} estimated with the 2018 \Plik\ likelihood (entries under the heading `\Plik' in Table \ref{tab:chi2_Planck_EDE}). The residuals with respect to the \Plik+SH$0$ES solution are shown by the orange lines in Fig.~\ref{fig:ede_spectra}
One can see that this solution is pulled by a small excess in power at  $\ell \gtrsim 1300$ in the \Plik\ TT and by an excess at  $\ell\sim 800-1000$ in the \Plik\ TE spectrum (see \cite{Smith:2022hwi} for discussion).  Neither of these features are strongly statistically significant and neither are seen in the higher signal-to-noise NPIPE spectra.

As reviewed in Ref.~\cite{Poulin:2023lkg}, in EDE models there has always been some residual tension between the CMB and the SH$0$ES distance scale.  However, if  EDE were the correct explanation of the Hubble tension,  we would expect to see a pull towards EDE as we  increase the signal-to-noise of the CMB spectra from \Plik\ to NPIPE. In fact, it is evident  from Fig. \ref{fig:ede_spectra}, that the NPIPE spectra move even closer to \LCDM\ without a hint of EDE.

Our conclusions on EDE models are  supported further by the profile likelihoods
of $f_{\rm EDE}$ and $H_0$ shown in  Fig.~\ref{fig:ede_profile} and reported in Table.~\ref{tab:EDE_profile}. For \Plik,  the profile likelihood allows for significantly larger values of $f_{\rm EDE}$ and $H_0$ than the posterior distribution extracted from the MCMC analysis, with a $\sim 2\sigma$ level preference for non-zero $f_{\rm EDE}$ \cite{Herold:2021ksg,Herold:2022iib}. The constraints derived from the NPIPE profile likelihood, although weaker than those determined from the posterior distribution, lead to the limits $f_{\rm EDE}<0.094$ and $H_0 < 70.2\, \Hunit$  at $2\sigma$. This indicates that for NPIPE,  the prior volume effects are minor and that there is no evidence for EDE from the CMB temperature and polarization spectra.

\begin{table*}[]
    \centering
    \begin{spacing}{1.4}
    \begin{tabular}{|c|c|c|c|c|}
    \hline
    & \multicolumn{2}{|c|}{NPIPE} & \multicolumn{2}{|c|}{\Plik} \\ 
    \hline
  SH0ES prior?  & no & yes & no & yes \\
\hline
 $h$
	 & $0.6811(0.684)^{+0.0047}_{-0.0082}$  
	 & $0.7124 (0.7167)\pm 0.0077$ 
	 & $0.6842 (0.6946)^{+0.0052}_{-0.011}$ 
	 & $0.7186 (0.7212)\pm 0.0078$ 
	 \\
$f_{\rm EDE}(z_c)$
	 & $ < 0.061(0.035)$ 
	 & $0.107(0.121)\pm 0.023$ 
	 & $ < 0.083 (0.064)$ 
	 & $0.128(0.135)^{+0.023}_{-0.021}$ 
	 \\
$\log_{10}(z_c)$
	 & $3.53(3.85)^{+0.27}_{-0.22}$ 
	 & $3.585(3.565)^{+0.049}_{-0.15}$ 
	 & $3.57(3.56)^{+0.25}_{-0.21}$ 
	 & $3.604(3.568)^{+0.014}_{-0.075}$ 
	 \\
$\theta_i$
	 & $1.91(3.01)^{+1.2}_{-0.62}$ 
	 & $2.53(2.82)^{+0.46}_{+0.079}$ 
	 & $1.93(2.76)^{+1.1}_{-0.72}$ 
	 & $2.73(2.76)^{+0.11}_{-0.090}$ 
	 \\
\hline
$\omega{}_{cdm }$
	 & $0.1216(0.1226)^{+0.0011}_{-0.0026}$ 
	 & $0.1300(0.1315)\pm 0.0029$ 
	 & $0.1229(0.1262)^{+0.0013}_{-0.0034}$ 
	 & $0.1329(0.134)\pm 0.0032$ 
	 \\
$10^{2}\omega{}_{b }$
	 & $2.226(2.233)\pm 0.016$ 
	 & $2.258(2.242)\pm 0.020$ 
	 & $2.253(2.253)^{+0.017}_{-0.022}$ 
	 & $2.282(2.274)\pm 0.022$ 
	 \\
$10^{9}A_{s }$
	 & $2.102(2.100)\pm 0.030$ 
	 & $2.140(2.140)\pm 0.031$ 
	 & $2.113(2.125)^{+0.029}_{-0.032}$ 
	 & $2.152(2.155)\pm 0.031$ 
	 \\
$n_{s }$
	 & $0.9691(0.9736)^{+0.0043}_{-0.0060}$ 
	 & $0.9868(0.9877)\pm 0.0059$ 
	 & $0.9713(0.9769)^{+0.0045}_{-0.0074}$ 
	 & $0.9902(0.9915)\pm 0.0059$ 
	 \\
$\tau{}_{reio }$
	 & $0.0543(0.0534)\pm 0.0070$ 
	 & $0.0561(0.0549)^{+0.0068}_{-0.0076}$ 
	 & $0.0549(0.0546)\pm 0.0070$ 
	 & $0.0561(0.0552)\pm 0.0073$ 
	 \\
\hline
$S_8$
	 & $0.830(0.831)\pm 0.011$ 
	 & $0.838(0.839)\pm 0.012$ 
	 & $0.835(0.841)\pm 0.011$ 
	 & $0.847(0.85)\pm 0.012$ 
	 \\
$\Omega{}_{m }$
	 & $0.3116(0.3111)\pm 0.0050$ 
	 & $0.3019(0.3009)\pm 0.0050$ 
	 & $0.3120(0.3096)\pm 0.0056$ 
	 & $0.3029(0.3027)\pm 0.0049$ 
	 \\

\hline
    \end{tabular}
\end{spacing}
    \caption{Credible interval and best fit values (in parentheses) in the EDE model reconstructed from analyses of the new NPIPE likelihood compared to the 2018 \Plik\ likelihood. We report results with and without the SH0ES prior on $M_b$. For parameters with two-sided constraints, we report the mean and   $1\sigma$ errors. For parameters with one-sided contraints, we report the $2\sigma$ limits. All of the MCMC chains include supplementary data  as described in the text.  }
    \label{tab:EDE_MCMC}
\end{table*}

\begin{table*}[]
    \centering
    \begin{tabular}{|c|c|c|c|c|}
   \hline
    & \multicolumn{2}{|c|}{NPIPE} & \multicolumn{2}{|c|}{\Plik}\\
     \hline
    & MCMC & Profile likelihood & MCMC & profile likelihood \\
\hline
 $h$
	 & $0.6811^{+0.0047}_{-0.0082}$ 
	 & $0.6837\pm0.0075$
	 & $0.6842^{+0.0052}_{-0.011}$ 
	 & $0.6944 \pm 0.0119$
	 \\
$f_{\rm EDE}(z_c)$
	 & $ < 0.061$ 
	 &  $0.035\pm 0.025 (<0.094)$
	 & $ < 0.083$ 
	 & $0.064\pm0.036 (<0.135)$
\\
\hline

    \end{tabular}
    \caption{(Bayesian) confidence intervals and (Frequentist) credible intervals extracted  from the MCMC posterior distribution and the likelihood profile respectively. Two-sided intervals are provided at 1$\sigma$, while one-sided  limits are provided at $2\sigma$.}
    \label{tab:EDE_profile}
\end{table*}

In conclusion, we have shown in this {\it letter} that a high signal-to-noise   likelihood constructed from the \Planck\ NPIPE maps precludes axion-like EDE from fully alleviating the Hubble tension, leaving a residual tension with the SH$0$ES distance scale  of about 3.7$\sigma$.  A comparison with the likelihood profile shows that our results are robust to prior-volume effects. Given the large parametric freedom of  the axion-like model studied in this paper, 
it seems unlikely that the Hubble tension can be solved by invoking an alternative EDE model (see Ref.~\cite{Poulin:2023lkg} for a review),  though we regard this possibility as a subject for future work. 

We note that Ref.~\cite{Goldstein:2023gnw} used low values of the spectral index $n_s$ inferred from observations of quasar  Ly$\alpha$ absorption lines to exclude the EDE models allowed by the \Plik\ likelihood. However, it is not  clear whether the matter power spectrum inferred from Ly$\alpha$ data is compatible with \Planck\ \cite{2020JCAP...04..038P} unless one also invokes a running of the spectral index (which would weaken the constraints on EDE). Various authors have derived tight  constraints on EDE models by including constraints on the shape of the matter power spectrum derived from galaxy surveys \cite[e.g][]{Hill:2020osr,DAmico:2020ods, Philcox:2022sgj}. However, such constraints are sensitive to model assumptions and choices of priors \cite{Herold:2021ksg, Smith:2020rxx,Simon:2022adh}.  The strong constraints on EDE reported in this paper are driven primarily by the linear CMB anisotropies and are therefore more robust than these analyses.

Our results exclude the $\sim 3\sigma$ preference for EDE reported by the ACT collaboration (with a preferred fraction $f_{\rm EDE} \sim 15\%$). The ACT DR4 best fit solution leads to a CMB temperature spectrum that is strongly excluded by both  \Plik\ and NPIPE. In fact, we have checked that one would need to discard the NPIPE temperature spectrum at $\ell > 1000$ to allow an EDE fraction as large as that favored by ACT DR4. There is no good reason to ignore these data. We therefore conclude that the ACT DR4 result is caused either by a systematic error in the temperature data, or a statistical fluke  (see also \cite{Smith:2023oop}).

 The absence of hint for physics beyond $\Lambda$CDM in the CMB anisotropies seems to argue against modifications of  pre-recombination physics as a solution of the Hubble tension.
Furthermore, it is well known that  late-time explanations are excluded by the combination of SN1a and Planck calibrated BAO mesurements \citep[see e.g.][for a recent discussion]{Brieden:2023}. In the absence of evidence in favour of EDE, and if  evidence  for the  Hubble tension persists as calibrations of the distance scale improve,  new physics at both early- and late-times  may be required to explain the discrepancy \cite{Vagnozzi:2023nrq}. Evidently,  a theoretical explanation of the Hubble tension remains elusive at the present time.

\begin{table*}[t!]
\def\arraystretch{1.2}
\scalebox{1}{
\begin{tabular}{|l|c|c|c|c|}
    \hline
         & \multicolumn{2}{|c|}{NPIPE} & \multicolumn{2}{|c|}{\Plik} \\

    \hline
    {\emph{Planck}}~high$-\ell$ TTTEEE  & 11237.4 & 11243.7& 2343.3 &2346.1 \\
    {\emph{Planck}}~low$-\ell$ TT & 22.0& 21.2& 22.1& 21.0\\
    {\emph{Planck}}~low$-\ell$ EE &395.9 &396.1 &396.1 &396.1  \\
    {\emph{Planck}}~lensing  & 9.5&10.1 & 9.4& 10.1\\
    BOSS BAO low$-z$ &1.2 & 1.9&1.3 & 1.7\\ 
    BOSS BAO/$f\sigma_8$ DR12 & 6.6&6.8 &6.8 &7.0 \\

    Pantheon+ &1411.3 & 1413.1&1411.5 &1412.8\\
    SH0ES & $-$ &4.4 &$-$  &2.5 \\
    \hline
    total $\chi^2_{\rm min}$ &13083.9 &13097.4 & 4190.4&4197.3\\
    $\Delta \chi^2_{\rm min}({\rm EDE}-\Lambda{\rm CDM})$ & -2.1 & -28.0 & -3.0 &-35.3 \\
    \hline
    $Q_{\rm DMAP}$&\multicolumn{2}{|c|}{3.7$\sigma$} &\multicolumn{2}{|c|}{2.6$\sigma$}\\
    \hline
\end{tabular}}
\caption{Best-fit $\chi^2$ per experiment (and total) for EDE when fit to the our baseline datasets, with either the Planck NPIPE or \Plik\ likelihood. We compare the fits with and without the SH$0$ES prior on $M_b$. We report the $\Delta \chi^2_{\rm min}\equiv\chi^2_{\rm min}({\rm EDE})-\chi^2_{\rm min}(\Lambda{\rm CDM})$ and the tension metric $Q_{\rm DMAP}$ defined in Eq. \ref{equ:Qdmap}.  Note that \emph{Planck} high-$\ell$ includes 10415 data points for NPIPE and 2289 bins for \Plik; the other data sets are the same in each column.}
\label{tab:chi2_Planck_EDE}
\end{table*}
\bibliographystyle{ieeetr}
\bibliography{biblio}
\begin{acknowledgements}
    The authors wish to thank Tanvi Karwal and Yashvi Patel for sharing their profile likelihood tool, Steven Gratton for his contributions to preparing NPIPE likelihoods and Marc Kamionkowski, Adam Riess, and Tristan L. Smith for valuable comments on the draft. GPE thanks the Leverhulme Trust for the award of an Emeritus Fellowship. VP is supported by funding from the European Research Council (ERC) under the European Union's HORIZON-ERC-2022 (Grant agreement No. 101076865).  The authors acknowledge the use of computational resources from the Excellence Initiative of Aix-Marseille University (A*MIDEX) of the Investissements d'Avenir program.  We acknowledge the use of LUPM's cloud computing infrastructure founded by Ocevu labex and France-Grilles. This project has received support from the European Union's Horizon 2020 research and innovation program under the Marie Skodowska-Curie grant agreement No.~860881-HIDDeN.
\end{acknowledgements}
\end{document}